\documentclass{elsart}

\usepackage{graphics,epsfig}

\usepackage{amssymb}

\def\lsim{\lower.5ex\hbox{$\; \buildrel < \over \sim \;$}}
\def\gsim{\lower.5ex\hbox{$\; \buildrel > \over \sim \;$}}

\def\t{\ifmmode {\tau} \else $\tau$ \fi}

\def\cm{\ifmmode {\rm cm}^{-1} \else cm$^{-1}$ \fi}
\def\s{\ifmmode {\rm s}^{-1} \else s$^{-1}$ \fi}
\def\cc{\ifmmode {\rm cm}^{-3} \else cm$^{-3}$ \fi}
\def\cs{\ifmmode {\rm cm}^{-2} \else cm$^{-2}$ \fi}
\def\g{\ifmmode \gamma \else $\gamma$\fi}
\def\l{\ifmmode \lambda \else $\lambda$\fi}

\def\t{\ifmmode \tau \else $\tau$\fi}
\def\G{\ifmmode \Gamma \else $\Gamma$\fi}
\def\Gt{\ifmmode \tilde{\Gamma} \else $\tilde{\Gamma}$\fi}

\def\kms{\ifmmode {\rm km\ s}^{-1} \else km s$^{-1}$\fi}

\def\ks{``knee"~}
\def\k{``knee"}

\begin{document}

\begin{frontmatter}



\title{Cosmic Ray Spectrum ``Knee": A Herald of New Physics?}


\author{D. Kazanas}

\address{NASA/GSFC, Code 661, Greenbelt, MD 20771}

\author{A. Nicolaidis}

\address{Department of Theoretical Physics, University of Thessaloniki,
Thessaloniki 54006, Greece}

\begin{abstract}
\noindent 
We argue that the knee  in the cosmic ray spectrum 
at energies $E \gsim 10^{15.5}$ eV is due to ``new physics",
namely to a channel in the high energy ($\gsim$ TeV in the CM) 
proton interactions hitherto unaccounted for in estimating the  
energies of the air shower cosmic rays. The new interaction 
transfers part of the primary particle's energy to modes which 
do not trigger the experimental arragement (neutrinos, lightest 
supersymmetric particle, gravitons ) thus underestimating its true
energy.  We show that this underestimate leads naturally  to the 
observed break (the ``knee") in the {\sl inferred} cosmic ray 
spectrum. The  suggestion we advance fits nicely to current theoretical 
extensions of the Standard Model (supersymmetry, technicolor, low 
scale gravity) where new  physics at the TeV scale manifests  
with the distinct signature of missing energy. We present a 
simple model where the new physics proceeds via gluon fusion 
and assuming a single power law for the galactic ($E \lsim 
10^{18.5}$ eV)  cosmic ray 
spectrum, we produce a good fit to the  data in the  $10^{14} - 
10^{18.5}$ eV range. Our proposal should be testable in 
laboratory experiments (LHC) 
in the near future and, should it be proven  correct, it would 
signal besides the presence of new physics in high  energy 
interactions, a drastically different interpretation of the  
sources and acceleration of cosmic rays.
\end{abstract}

\end{frontmatter}
\label{intro}
The origin of cosmic rays is a subject which, despite the observational
and theoretical progress made since their discovery has not been 
settled as yet. The reason can be traced to the breadth 
of their spectrum which extends over 11 orders of magnitude to 
$\gsim 10^{20}$ eV (see e.g. \cite{gaiss00} for a recent review) and 
the fact that by virtue of their diffusion through the galaxy 
most information concerning their sources is practically lost.
Thus, to zeroth order, especially at higher energies ($> 10^{12}$
eV) at which cosmic ray composition measurements are difficult, 
the sole source of clues about the cosmic ray origin and acceleration
is their over all spectrum.

The cosmic ray spectrum consists, roughly speaking, of three 
distinct sections, each of power law form, $E^{-\g}$ in the particle 
energy $E$, but with different values for the index \g: 
In the range $10^9 - 10^{15.5}$ eV the index $\g \simeq 2.75$. 
Above this energy (the \k), the spectrum steepens  to a power law of
$\gamma \simeq 3$ which extends to $\sim 10^{18}$ eV, with 
some evidence for a further steepening in the spectrum
indicating a possible cut-off at $E \simeq 10^{18.5}$ eV
\cite{bird93}. 
This steepening is reversed  at slightly larger energies (at the 
``ankle") with the spectrum flattening to $\g \sim 2-2.5$ and 
extending to $E \simeq 10^{20.5}$ eV, at which point the existing 
statistics are too poor to provide a well defined flux measurement. 

Considering that cosmic rays propagate in the galaxy by diffusion 
through the tangled interstellar magnetic field, one can argue 
convincingly that particles with gyroradii larger than the galactic 
scale height ($\sim 1$ kpc) ought to be extragalactic. Given 
that the gyroradius of a proton of energy $E(eV)$ is $R_g 
\sim 1 {\rm kpc}\; E_{18} /B_{-6}$ (where $E_{18} = E({\rm eV})
/10^{18}$ and $B_{-6}$ is the galactic magnetic field in 
$\mu G$), it is expected that protons of energy $E \gsim
10^{18.5}$ would escape freely from the galaxy. This notion 
is in agreement with the indication of an additional steepening or
a potential cut-off in the spectrum at $E \gsim 10^{18}$ eV, 
with the subsequent flattening  at higher energies being naturally 
interpreted as due to a ``harder" extragalactic component.

This latter extragalactic component, which includes several 
events above $10^{20}$ eV, has caught recently the attention of 
the community: It is well known \cite{greis66} that protons of 
energies $\gsim 10^{19.5}$ eV suffer catastrophic photopion production 
losses on the Cosmic Microwave Background (CMB). If this 
extragalactic component permeates uniformly all space, as it was 
thought to be the case, this process should then lead to a cut-off 
(the so-called GZK cut-off) rather than an excess flux above this 
energy. The potential identification of the source of this component 
with either gamma ray bursts within 100 Mpc \cite{Wax96}, a novel, 
neutral hadron immune to the photopion losses \cite{farbier98}, 
or the decay of heavier Big--Bang relics \cite{hillsch} has provided 
the impetus for a recent flare of activity regarding the origin 
of this specific part of the cosmic ray spectrum (see \cite{stecker}
for a review). 

However, it is not only this highest energy regime which defies 
our understanding  of the cosmic ray spectrum origin. The spectrum 
at energies $E \lsim 10^{18.5}$ eV, thought to be galactic (see 
however \cite{prosz92}) and presumably easier to comprehend, 
challenges on its own right our understanding of its origin, 
arguably more severely than the extragalactic component, 
at least from the purely ``fitting the curve" point of view.

This last statement, while counter intuitive at first glance, 
it is nonetheless true, as shown by the following simple argument (see 
also \cite{Ax91}): It is easy to obtain a ``flattening" of the 
(any) spectrum by combining two independent components, since 
the harder one  will always dominate at sufficiently high energies.  
This appears to be the case with the cosmic ray spectrum at $E \gsim
10^{18.5}$ eV. 
On the other hand, producing a continuous spectrum with a steepening 
break similar to that observed at the cosmic ray spectrum \ks 
is much harder: It demands the presence of {\sl two distinct} 
acceleration mechanisms, one of which carries the particles 
to the \ks with spectrum $\propto E^{-2.75}$ and a second one 
which takes practically {\sl all} the particles that reach the 
\ks via the first mechanism {\sl and only these}, to a thousandfold 
higher energy with spectrum $\simeq E^{-3}$. If this second acceleration 
mechanism accelerated only a fraction of the particles that reach 
the \ks (a perfectly ``reasonable" assumption for most acceleration
processes), it would lead to a (not observed) 
discontinuity in the spectrum at this energy. 

To complicate matters
further, the most promising acceleration mechanism of galactic
cosmic rays, namely supernova shocks, can barely produce 
(even theoretically) particles of energies as high as $E_M \simeq
Z \cdot 10^{14}$ eV \cite{lagces83} (the nuclear charge $Z$ appears 
because of the dependence of the gyroradius on it), 
even with the diffusion coefficient 
at the Bohm value \cite{bervolk00}. Energies as high as that of the
\ks can be achieved only by assuming that the cosmic ray composition 
at this point consists mainly of Fe nuclei. There exists no known 
(to the authors) mechanism which would carry even a fraction of 
the (diffusing through interstellar space) particles of the \ks to 
the energy of the ``ankle", in a way that produces the observed 
spectrum. Furthermore, it is hard to argue that the transition 
between the two power laws at the \ks could be due to some unknown 
transport effect of the cosmic rays in the galaxy, because it appears
to be much too sharp for such an interpretation.

Motivated by the above considerations we are led to propose that
the break at the \ks of the cosmic ray spectrum is indicative, 
not of a distinct acceleration mechanism, but of the emergence of 
``new physics" in the high energy proton interactions, namely of 
a new channel beyond those considered in the models employed to 
infer the primary particle energy in the air shower arrays. We argue
that if a fraction of the energy associated with this new 
channel is in a form that does not trigger these detectors, it will 
result to an underestimate of the primary particle's energy (a 
similar suggestion was made in \cite{nikol93}, based on the 
interpretation of certain peculiarities in the Extensive Air
Shower (EAS) data
as a sharp change in inelasticity with energy). For a cosmic 
ray spectrum which is a {\it single} power law in energy, this 
underestimate will then manifest as an increase in its slope (a \k) 
at the energy at which this new channel turns-on, with the spectrum 
reverting to its original slope when it eventually saturates. 
Furthermore, to account for the break observed at the \ks of the 
cosmic ray spectrum, this new  channel should ``turn-on" at an energy 
$\simeq$ TeV at the  center of mass, a scale tantalizingly close to 
that at which the emergence of ``new physics" is anticipated on the 
basis of rather general considerations \cite{lee77}.

As candidates of such new physics we cite supersymmetry (SUSY), a
symmetry that interrelates bosons and fermions \cite{S1ref}, 
technicolor, the postulate that the Higgs boson is composite made
of ``techni-fermions" much the way that the proton is made of quarks
\cite{Tcref}, or the postulate that the observed Planck scale 
$M_P \simeq 10^{19}$ GeV is a $4-$dimensional epiphenomenon of the 
``true" scale of gravity which is of order of 1 TeV,  ``living"
in a higher dimensional world of $D = n+4$ dimensions \cite{X1ref}. According 
to this last proposal, the gravitational interaction would acquire 
the strength of the rest interactions at the characteristic scale
of $\sim 1$ TeV, with prolific  graviton production (which of course
do not register in the EAS detectors) in particle interactions 
above this energy. Interestingly, energy which does not register in 
the EAS detectors (needed to account for the  break at the \k) is 
predicted also by the other two proposals: In SUSY this is the 
lightest supersymmetric particle (LSP). In techni-color models, 
$p\,p$ collisions at energies $\gsim 1$ TeV produce techni-hadrons
(hadrons carrying techni-color besides the ordinary color); 
techni-hadron decay leads to $W'$s of which a significant fraction decays 
in neutrinos that also do not trigger the EAS detectors.

One should note from the outset that EAS experiments, by observing the 
results of reprocessing high energy interactions of a largely unknown 
composition through 1000 gm cm$^{-2}$ of atmosphere, can provide but 
the most general characteristics of these events and as such are not
a substitute for the more detailed accelerator experiments. However, 
some of the general properties of these interaction can be deduced
in these experiments. Such an example is the logarithmic
increase of the strong interaction cross section with energy, which 
was first inferred from the energy dependence of the height of 
maximum development of cosmic ray showers. The arguments on the 
nature of the break at the \ks (i.e. that it is convex rather than concave)
are of this general character and their interpretation should be viewed 
in the same spirit.
 
To avoid making a specific choice from the list of available 
alternatives at this early stage of our investigation, we model 
the process simply as the production and 
decay of a system of total invariant mass $M_0 = 2$ TeV and we 
parameterize the entire process by two parameters: the fraction $y$ of 
the primary particle's energy that registers in the cosmic ray 
detectors and the asymptotic (i.e. at energies much higher than the 
production threshold) ratio $\alpha$ of the cross section 
associated with this new channel to that of the standard interactions.

On dimensional grounds, the cross section of the new channel is 
assumed to be of the form 
\begin{equation}
\sigma_n(E) = \frac{B}{s}g(\tau) 
\label{xsec}
\end{equation}
where $B$ is a dimensionless constant (related to $\alpha$), 
$s = 2 m_p E$ and $g(\tau)$ is a function of the dimensionless 
ratio $\tau = M_0^2/s$. At high energies we expect the $p \, p$
interactions to be dominated by gluon scattering and accordingly
\begin{equation}
g(\tau)=  \int_{\tau}^1 f(x) f(\tau/x) \; dx
\label{gt}
\end{equation}
where $f(x)$ is the gluon distribution within the proton, which 
we parameterize as
\begin{equation}
f(x) = \frac{1}{2}(N+1) \frac{(1-x)^N}{x}~,~ ~{\rm with}~~~N~= ~6
\label{gludis}
\end{equation}

For the conventional $p\,p$ interactions the cross section 
rises slowly with energy and for our purposes we consider it
to be a constant $\sigma_o(E) \simeq 80$ mbarn. At energies
above the new physics threshold the cosmic ray interactions
will proceed either through the standard channels with 
probability $P_o(E) = \sigma_o(E)/[\sigma_o(E)+\sigma_n(E)]$ 
or through the new channel with probability $P_n(E) = \sigma_n(E)/
[\sigma_o(E)+\sigma_n(E)]$; given that $\tau g(\tau) \rightarrow 1/C_6 =$ 
constant for $\tau \rightarrow 0$, setting $B = M_0^2 C_6 \sigma_o \alpha$
we get $\sigma_n(E)/\sigma_o(E) \rightarrow \alpha$ for $E \gg M_0^2/2m_p$ 
as desired. 

Whenever high energy cosmic ray particles interact through the new
channel, events of total energy $E^{\prime}$ will register at 
the detector an energy $E = y  E^{\prime} ~(y<1)$. Therefore, 
if the cosmic ray intensity is $I(E)$, for particles interacting 
through this new channel, 
the {\sl inferred} intensity will be of the form
\begin{equation}
\int I(E^{\prime}) P_n(E^{\prime}) \delta(E - y E^{\prime}) \, 
dE^{\prime} = \frac{1}{y}\; I\left(\frac{E}{y}\right) 
P_n\left(\frac{E}{y}\right)
\label{newcrflu}
\end{equation}
while for events interacting through the conventional channel the 
resulting intensity will have a similar form but with $y=1$ and 
$P_o(E)$ in place of $P_n(E/y)$.

Assuming the incident {\sl galactic} cosmic ray spectrum to be of the form 
$I_{IG}(E) = E^{-\g}exp(-E/E_0)$  with $\g \simeq 2.75$ and $E_0 \simeq 
10^{18.5}$ eV, a value consistent with the earlier argument on the 
cosmic ray gyroradii at $E \simeq E_0$, the observed cosmic 
ray flux at an energy $E$ will be
\begin{equation}
I_O(E) = E^{-\gamma} e^{-E/E_0} \Big[ \frac{1}{1 + C_6\alpha\,\tau 
g(\tau)} 
+ \frac{\alpha \; y^{\g-1}\; e^{-E / yE_0 + E/E_0}} 
{ 1 /[C_6 (\tau y) g(\tau y)]+ \alpha} \Big]
\label{crfull1}
\end{equation}

The first term in the square brackets in Eq. (\ref{crfull1})
represents the contribution to the spectrum from interactions
through the conventional channels while the second that due to the 
new one. The presence of the exponential cut-off broadens and 
deepens the effects of the presence of the new channel. Their 
combined effect is necessary for a good fit to the data. 

\begin{figure}[hbt]
\centerline{\epsfig
{file=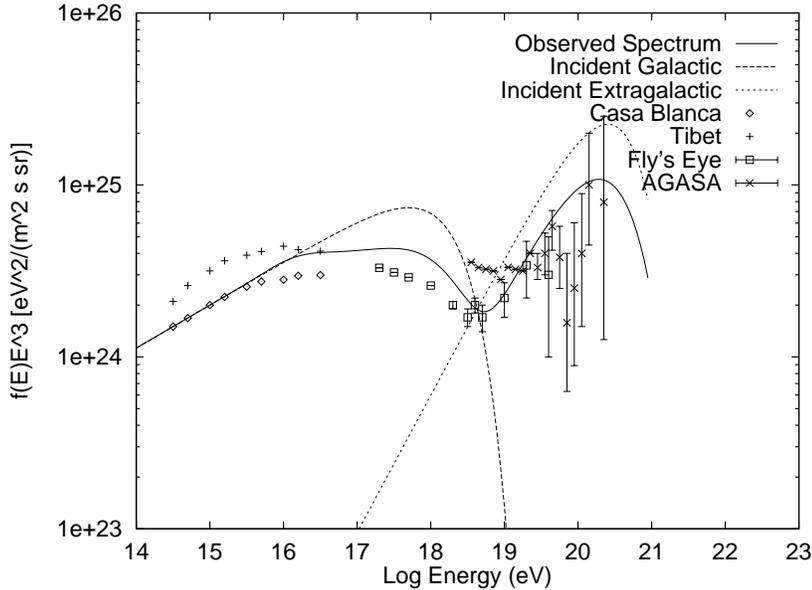,width=.8\textwidth,angle=0}}
\caption{The cosmic ray spectrum $f(E)$ multiplied by $E^3$ for 
$E>10^{14}$ eV. Long and
short dashed lines are respectively the incident galactic and 
extragalactic components. Solid line is the combined spectrum expected 
to be measured for $y=1/2,~\alpha=2$.
$+$'s are the Tibet data, diamonds the Casa Blanca data, squares the 
Fly's Eye data and $\times$'s the AGASA data.}
\label{f2}
\end{figure}

In addition to the galactic cosmic ray component, whose contribution 
is expected to be unimportant beyond $E_0 \simeq 10^{18} {\rm eV}$, 
there exists also an extragalactic component, whose reprocessing
in the atmosphere should also result in a modification of its 
spectrum according to the prescription of Eq. (\ref{crfull1}).
The precise form of this component is of course unknown since it
is dominated at lower energies by the galactic component. Following
\cite{gaiss00} we assume its spectrum to be of the form $I_{EG}(E) 
\propto E^{-q} exp(-E/E_1)$ with $q \simeq 2.2 - 2.6$ and $E_1 
\simeq 10^{20}-10^{20.3}$ eV.  In Figure \ref{f2} we present the 
entire (galactic + extragalactic) cosmic ray spectrum from $10^{14}
- 10^{21}$ eV, with $\g = 2.75, q=2.2, E_0 = 2\;10^{18} {\rm eV}, E_1
= 10^{20.3} {\rm eV}$ by applying the effects of the new postulated 
channel in the interaction to both components with $y=1/2,~\alpha =2$. 
We also plot the 
relevant data from two different experiments in each of the $10^{14} - 
10^{16}$ eV (Tibet \cite{tibet}, CASA-BLANCA \cite{casab}) 
and $10^{18} - 10^{20}$ eV (Fly's Eye Stereo \cite{bird93}, AGASA 
\cite{agasa}) energy ranges. We expect that these should
bracket the true values of the corresponding parameters and 
should serve as a gauge of the systematic errors involved 
in computing the cosmic ray spectra in each range. 

While it is very difficult to draw immediate conclusions 
favoring specific models from the existing data our general 
considerations appear to be on the correct footing: Our 
proposal produces a break in a single power law cosmic ray
spectrum of the correct character (an increase in the slope)
over a very limited range in energy (half a decade), as demanded
by the data. As noted earlier,  both these features are very
hard to produce with more conventional approaches. Our proposal 
is further supported by the more detailed analysis of EAS data:
the apparent 
very sharp change in cosmic ray composition to almost exclusively 
Fe, inferred from the abrupt change in the depth of the maximum in 
the shower development around $10^{16}$ eV (fig.5 of \cite{casab}), 
is qualitatively of the form  expected by a sharp increase in the 
interaction  cross section, such as we propose, and an ensuing 
dispersion of the available energy to a large number of secondary 
particles. Such an abrupt change in composition is very hard to 
understand (let alone model) within the realm of conventional models.

Considering the fit of our figure at energies higher than the \k, one
notes that the assumption of a cut-off in the galactic component at 
$\sim 19^{18.5}$ eV is necessary for a good fit, but it is also 
reasonable, supported both theoretically (the gyroradii arguments
above) and experimentally (as discussed in \cite{gaiss00}) by the
observed increase in the cosmic ray anisotropy at this energy
\cite{agasa}. Given the simplicity of our assumptions we think that
this fit is particularly good. One could think of several ways of
improving it, if necessary, at the expense of introducing additional
parameters in modeling the physics of the high energy interactions.
We think that the present quality of the data does not warrant 
at this point such an extension of the assumptions used. The presence
of the extragalactic component does affect the values of the 
fitting parameters since this component does make some contributions 
at lower energies too (the values of $q$ and $E_0$ are thus closely 
related). However, this seems to affect little the values of $y$ and
$\alpha$ which determine primarily the fit of the spectrum at the \k.

Where all these leave us? Our interpretation carries with it a 
number of consequences: (a) To start with, it implies the presence 
of ``new" structure in the high energy physics interactions at 
energies consistent with those suspected on the basis of 
generic theoretical considerations. The new physics is slightly 
beyond the reach of the Fermilab Tevatron, but it will be preeminently
present at LHC. Their ``benchmark"  signature, based on our 
interpretation of the \k, should be a new
channel with strong  interaction cross section and missing
energy. We have computed the cross-section of the new 
channel at Tevatron energies using the formalism presented above 
and it was found to be a small fraction of the total cross section.
At this point we would like to refrain from more specific predictions
given the simplicity of our model and the quality of the cosmic
ray data; however, the above requirements are quite strong and may 
already be in disagreement with some of the above proposed alternatives.
 (b) On the cosmic ray physics side, the demand for a break at 
the \ks much sharper than allowed by cosmic ray transport 
coupled with our proposal and fits, lead to
the radical suggestion that the cosmic ray sources must, by and
large, produce spectra extending to the ``ankle" (rather than the \k). 
This then leads to the unsettling conclusion that supernovae should
not be {\sl the} dominant contributor to the cosmic ray spectrum
at least at energies greater than $\sim 10^{14} - 10^{15}$ eV.
Leaving for the moment the issue of cosmic ray composition aside, 
our interpretation of the cosmic ray spectrum at the \k, implies 
the presence of an alternate source of cosmic rays capable of 
accelerating particles to the energies of the ``ankle".
It is  interesting to note that
independent considerations recently pointed to similar conclusions
\cite{dermer}. Hints to the nature of these sources may in fact be 
provided by the observed anisotropy at $E \sim 10^{18}$ eV toward 
the galactic center \cite{agasa1}. We plan to revisit both
these issues in a future publication.

While this paper was being written, the potential effects of physics
beyond the Standard Model were announced (deviation of the muon $g-2$
value from that of the standard model \cite{muong}). This effect 
was interpreted as requiring the presence of new physics at $\sim$ 
TeV scales, similar to that involved in our considerations.

DK  would like to thank Frank Jones, John Krizmanic and Bob 
Streitmatter for a number of discussions, comments, criticism 
and encouragement, as well as Greg Thornton and Francis Halzen for
very useful correspondence and constructive criticism 
in relation to specific issues of the paper. AN would like to 
acknowledge useful discussions 
with participants of the Erice meeting  ``Phase Transitions in 
the Early Universe" in particular Floyd Stecker and Peter Biermann
and thank the organizers for an interesting and 
stimulating meeting. AN was partially supported by  the EU TMR 
programme, contract  FMRX-CT98-0194.


\begin{thebibliography}{00}




\bibitem{gaiss00}
T. K. Gaisser, astro-ph/0011524

\bibitem{bird93}
D. J. Bird, {\it et al.}, Phys. Rev. Lett. {bf 71}, 3401 (1993)

\bibitem{greis66} K. Greissen, Phys. Rev. Lett. {\bf 16}, 748 (1966);
G.T. Zatsepin and V. A. Kuzmin, Pis'ma Zh. Eksp. Th. Fiz., {\bf 4}, 
114 (1966)[JETP Lett. {\bf 4}, 78 (1966)]; 

\bibitem{Wax96}
E. Waxman, Phys. Rev. Lett. {\bf 75}, 386 (1995); M. Vietri, 
Astrophys. J. {\bf 453} 883 (1995)

\bibitem{farbier98}
G. R. Farrar and P. L. Biermann, Phys. Rev. Lett. {\bf 81}, 3579 (1998)

\bibitem{hillsch}
C. Hill, D. N. Schramm, and T. P. Walker, Phys. Rev {\bf D 36}, 1007 (1987)

\bibitem{stecker}
F. W. Stecker, astro-ph/0101072


\bibitem{prosz92}
R. J. Protheroe and A. P. Szabo, Phys. Rev. Lett. {\bf 69}, 2885 (1992)

\bibitem{Ax91}
I. Axford, {\it Proc. Particle Acceleration in Cosmic Plasmas.}, AIP Conf. 
Proc. No 264, G. P. Zank, T. G. Gaisser eds., p. 45 (1991) 

\bibitem{lagces83}
P. O. Lagage and C. J. Cesarsky, Astron. Astrophys. {\bf 118} 223 (1983)

\bibitem{bervolk00}
E.G. Berezhko  and H.J. V\"olk, Astron. Astrophys. {\bf 357} 283 (2000)


\bibitem{lee77}
B. W. Lee, C. Quigg, H. Thacker, Phys. Rev. {\bf D 16} 1519 (1977)

\bibitem{nikol93}
S. I. Nikolsky, Bull. Russ. Acad. Sci. {\bf 57} 595 (1993)

\bibitem{S1ref}  
R. Mohapatra, Unification and Supersymmetry, Springer - Verlag (1986)

\bibitem{Tcref}
K. Lane, hep-ph/0006143 (and references therein).

\bibitem{X1ref} 
I.  Antoniadis, Phys. Lett. B 246, 377 (1990);
N. Arkani-Hamed, S. Dimopoulos, and G. Dvali, Phys. Lett. B 429, 263 (1998);
I. Antoniadis, N. Arkani-Hamed, S. Dimopoulos, and G. Dvali, Phys. Lett. B 436, 257 (1998) 


\bibitem{tibet}
M. Amenomori {\it et al.} Astropart. Phys. {\bf 10} 137 (2000)

\bibitem{casab}
J. W. Fowler {\it et al.} astro-ph/0003190 (submitted to Astroparticle Physics)

\bibitem{agasa}
M. Takeda {\it et al.} Phys. Rev. Lett. {\bf 81} 1163 (2000)

\bibitem{agasa1}
N. Hayashida {\it et al.} Astroparticle Physis {\bf 10} 303 (1999)



\bibitem{dermer}
C. D. Dermer, in {\it Heidelberg $\gamma$ 2000}, eds. F. A. Aharonian and
H. V\"olk (AIP: New York) in press (astro-ph/0010564)

\bibitem{muong}
H. N. Brown {\it et al.} Phys. Rev. Lett. {\bf 86} 2227 (2001)






\end{thebibliography}
\end{document}